\documentclass[journal,twoside,web]{ieeecolor}
\usepackage{generic}
\usepackage{amsmath,amssymb,amsfonts}
\usepackage{graphicx}
\usepackage{algorithm}
\usepackage{algorithmic}
\usepackage{hyperref}
\usepackage{textcomp}
\usepackage{booktabs}
\usepackage{makecell}
\usepackage{pifont}
\usepackage{multirow}

\begin{document}

\title{CA-TCN: A Causal-Anticausal Temporal Convolutional Network for Direct Auditory Attention Decoding}
\author{I. Garcia-Ugarte, R. Eguinoa, R. San Martin, D. Paternain, \IEEEmembership{Member, IEEE}, and C. Vidaurre
\thanks{This work has been submitted to the IEEE for possible publication. Copyright may be transferred without notice, after which this version may no longer be accessible.}
\thanks{Grant PID2020-118829RB-I00 supported IGU, CV, RE and RSM,  grant PID2024-161502OB-I00 supported IGU, CV, RE and RSM, grant PREP2024-003050 supported IGU, grant PRE2021-099863 supported RE, BERC 2022-2025 program of the Basque Government, BCBL Severo Ochoa excellence accreditation CEX2020-001010/AEI/10.13039/501100011033 and IKERBASQUE supported CV.}
\thanks{This work involved human subjects in its research. Approval
of experimental procedures and protocols was granted by
the Ethics, Animal Experimentation and Biosafety Committee of the Public University of Navarra with number PI-029-20\_CV.}
\thanks{I. Garcia-Ugarte was with the Department of Science, Universidad Pública de Navarra (UPNA), 31006 Pamplona, Spain, and currently is with BCBL, Basque Center on Cognition Brain and Language, 20009
San Sebastián, Spain (e-mail: inigo.garciaugarte@unavarra.es).}
\thanks{R. Eguinoa and R. San Martin are with the Department of Science, Universidad Pública de Navarra (UPNA), 31006 Pamplona, Spain (e-mail: ruben.eguinoa@unavarra.es; ricardo.sanmartin@unavarra.es).}
\thanks{D. Paternain is with the Department of Statistics, Computer Sciences and Mathematics, Universidad Pública de Navarra (UPNA), 31006 Pamplona, Spain. (e-mail: daniel.paternain@unavarra.es).}
\thanks{C. Vidaurre is with Ikerbasque, Basque Foundation for Science, 48009 Bilbao, Spain, with BCBL, Basque Center on Cognition Brain and Language, 20009
San Sebastián, Spain, and also with the Machine Learning Group, Technische Universitat Berlin, 10587 Berlin, Germany (e-mail:
cvidaurre@bcbl.eu).}}

\maketitle

\begin{abstract}

A promising approach for steering auditory attention in complex listening environments relies on Auditory Attention Decoding (AAD), which aim to identify the attended speech stream in a multiple speaker scenario from neural recordings. Entrainment-based AAD approaches, typically assume access to clean speech sources and electroencephalography (EEG) signals to exploit low-frequency correlations between the neural response and the attended stimulus. In this study, we propose CA-TCN, a Causal–Anticausal Temporal Convolutional Network that directly classifies the attended speaker. The proposed architecture integrates several best practices from convolutional neural networks in sequence processing tasks, including residual connections, depthwise convolutions, and dilated filters. Importantly, it explicitly aligns auditory stimuli and neural responses by employing separate causal and anticausal convolutions respectively, with distinct receptive fields operating in opposite temporal directions. Experimental results, obtained through comparisons with three baseline AAD models, demonstrated that CA-TCN consistently improved decoding accuracy across datasets and decision windows, with gains ranging from 0.5\% to 3.2\% for subject-independent models and from 0.8\% to 2.9\% for subject-specific models compared with the next best-performing model, AADNet. Beyond accuracy, the model complexity and real-time constraints were also studied. Moreover, CA-TCN demonstrated spatial robustness across different datasets with initial EEG spatial filters sharing a similar weight distribution. Overall, this work introduces an accurate and unified AAD model that outperforms existing methods while considering practical benefits for online processing scenarios. These findings contribute to advancing the state of AAD and its applicability in real-world systems.

\end{abstract}


\section{Introduction}
\label{sec:introduction}
\IEEEPARstart{I}{n} complex auditory environments, such as the well-known Cocktail Party scenario ~\cite{cherry_experiments_1953}, multiple sound sources compete simultaneously for perceptual dominance. In these situations, the human auditory system is able to selectively attend to a target source while suppressing irrelevant acoustic information. For individuals with hearing impairment, however, this selective filtering mechanism is often severely compromised, leading to significant difficulties in everyday listening conditions. Conventional hearing aids frequently fail to provide effective assistance in such scenarios, which is reflected in their limited adoption, with only 10–20\% of hearing-impaired individuals using them on a regular basis~\cite{lesica_why_2018}. Recent advances in deep learning–based sound source separation algorithms~\cite{luo_convtasnet_2019, wang_deep_2017} offer the potential to extract clean speech signals from complex acoustic mixtures. Nevertheless, a key challenge remains: identifying and selecting the attended source in a seamless and user-driven manner~\cite{hjortkjaer_realtime_2025}. In this context, auditory attention decoding (AAD) aims to detect attended source directly from neural activity, emerging as a promising cognitively driven solution that could enhance hearing aids performance~\cite{geirnaert_electroencephalographybased_2021}. 

Numerous studies have demonstrated the feasibility of decoding auditory attention including electroencephalography (EEG) \cite{aiken_human_2008, osullivan_attentional_2015}, magnetoencephalography (MEG) \cite{akram_robust_2016} and electrocorticography (ECoG) \cite{dijkstra_identifying_2015, raghavan_improving_2024} measures. Among these, EEG is considered the most suitable for real-world applications due to its scalability, wearability, and relatively low cost. Many cognitive processes associated with attention are reflected in neural oscillations \cite{meyer_neural_2018}, several of which have been exploited to identify the attended sound source. 


When the auditory signals are available, cortical entrainment-based AAD approaches, which typically rely on correlation measures to identify the attended speech stream, have been shown to be robust and to generalize well across trials and subjects \cite{geirnaert_electroencephalographybased_2021, nguyen_aadnet_2025}. Moreover, these algorithms have already been implemented and validated in closed-loop online AAD scenarios \cite{aroudi_closedloop_2021, hjortkjaer_realtime_2025}. Since auditory stimuli are available in the present study, we focus on entrainment-based AAD algorithms. These methods leverage the so-called entrainment property, where a sequence of acoustic cues drives the cycles of a given neural oscillation into a phase aligned rhythmic sequence \cite{meyer_synchronous_2020}. This low-level neural mechanism is elicited on EEG signals at low-frequency bands, typically Delta (1-4 Hz) and Theta (4–8 Hz) \cite{viswanathan_electroencephalographic_2019}, where a phase-locking phenomenon occurs between the envelope of the speech and neural oscillations. 

AAD studies typically address the identification of the attended source in a two-speaker competing scenario, in which two main modeling paradigms can be distinguished. Forward models aim to predict the evoked EEG responses from the acoustic stimuli, whereas backward models reconstruct stimulus representations directly from the EEG signals \cite{wong_comparison_2018}. In practice, backward approaches are more commonly adopted, as they provide better performance~\cite{geirnaert_electroencephalographybased_2021}. Backward models can be further categorized into linear and non-linear approaches. Linear backward models attempt to reconstruct the attended speech envelope by applying a set of linear filters to time-lagged EEG signals \cite{osullivan_attentional_2015, wong_comparison_2018}. Then reconstructed envelope is correlated with ground-truth speech envelopes and the stimulus yielding the highest Pearson’s correlation coefficient is identified as the attended one. Alternatively, in \cite{decheveigne_decoding_2018}, a hybrid linear approach combining forward and backward models through Canonical Correlation Analysis (CCA) was proposed. 

More recently, due to the progress achieved across data processing fields using deep learning techniques, several non-linear models have been proposed to replace linear backward regression algorithms for AAD. More recently, advances in deep learning across multiple data processing domains have motivated the development of non-linear models to replace traditional linear backward regression approaches for AAD. For example, in \cite{thornton_robust_2022}, two deep learning-based backward models were proposed to reconstruct the attended speech envelope: one relying on a Fully Connected Neural Network (FCNN) and another based on a Convolutional Neural Network (CNN).

All previous mentioned entrainment-based AAD algorithms perform attended speech selection in a two-step process. First, the AAD model is applied to the EEG signal (or both EEG and stimulus in the case of CCA), and then the resulting predicted envelope or temporal sequence classification. The Auditory Attention Decoding Network (AADNet) \cite{nguyen_aadnet_2025} proposes a deep-learning based architecture that unifies these two steps to directly identify the attended stimulus. As a result, the whole network is optimized based on the final classification accuracy. Importantly, the results were obtained under a clear validation scheme using publicly available data, which prevent the evaluation setting from being biased by some common overfitting issues in AAD  ~\cite{rotaru_what_2024, puffay_relating_2023}. Specially for the subject-independent validation scheme, where no user specific data is used during training, AADNet demonstrated a substantial improvement in accuracy when compared with the other established AAD models. The architecture of AADNet is inspired by the Inception network paradigm~\cite{szegedy_going_2015}, which relies on multiple parallel convolutional branches applying kernels of different sizes to capture diverse characteristics of the input signals. While effective, this design requires the careful specification of a large number of branch-specific hyper-parameters, thereby constraining each branch to model a predefined temporal or spatial feature. From our perspective, several established design choices in modern convolutional network design~\cite{zhang_modern_2021}, i.e., residual connections~\cite{he_deep_2016} and dilated convolutions for temporal sequence modeling~\cite{bai_empirical_2018a, lea_temporal_2017}, offer the opportunity to simplify the architecture while promoting a more unified and robust modeling strategy for AAD.

Motivated by these considerations, this work introduces CA-TCN, a Causal–Anticausal Temporal Convolutional Network designed to directly classify the attended speaker in a two-speaker competing scenario. The proposed architecture builds upon a Temporal Convolutional Network (TCN) framework to jointly project and align the signal representations differentiating the processing of the two types of signals. In fact, stimuli is processed by causal convolutions considering a near one-second long past receptive field whereas EEG signals are processed in an anticausal fashion including a tighter receptive field. An ablation study was performed to further discuss and study the incorporation of these design choices as well as the contribution of each processing stage in the overall accuracy. When compared with other three established AAD models, CA-TCN outperformed them consistently across the three datasets and the two validations schemes considered in the study. Moreover, the clustering performed on the spatial filters of the EEG branch, highlighted the stability of the spatial processing across different sets. 

We argue that CA-TCN represents an important step toward a viable AAD real-world system, as it provides an accurate and efficient approach for entrainment-based AAD.

\section{Materials and Methods}

\subsection{Baseline Auditory Attention Decoding models}
\label{sec:aad_algorithms_review}

\subsubsection{Linear backward models (Ridge regression)}

Linear backward models \cite{osullivan_attentional_2015} aim to reconstruct the attended envelope ($\hat{s_a}$) by applying a set of linear filters ($d \in \mathbb{R}^{1\times LC}$) to a lagged version of the recorded EEG signal ($\mathbf{X} \in \mathbb{R}^{T\times LC}$):
\begin{equation}
    {\hat{s}_a(t)} = \sum^C_{c=1}\sum^{L-1}_{l=0}d_c(l)x_c(t+l)
    \label{eq:ridge}
\end{equation}
Where $C$ denotes the number of EEG channels and $L$ the total number of time-lags.
The matrix $\mathbf{X}$ is formed by concatenating $L$ lagged versions of the EEG signal. 


The decoder matrix is obtained by minimizing the mean squared error (MSE) between the estimated and true attended envelopes, leading to the following closed-form solution:
\begin{equation}
    d=(\mathbf{X^T X} + \mathbf{\lambda} \mathbf{I} )^{-1}\mathbf{X^T}s_a 
    \label{eq:ridge_solution}
\end{equation}
where an L2 regularization, controlled by the parameter $\lambda$, is included to reduce overfitting  \cite{wong_comparison_2018}. Here, $\mathbf{X}^\mathrm{T}\mathbf{X}$ corresponds to the auto-correlation matrix, and $\mathbf{X}^\mathrm{T}s_a$ to the cross-correlation matrix.
The envelope that yields the higher correlation coefficient is identified as the attended stimulus.

\subsubsection{Canonical Correlation Analysis (CCA)}

The Canonical Correlation Analysis (CCA) method is designed to maximize the correlation between two multivariate domains. When applied to AAD \cite{decheveigne_decoding_2018}, a set of decoder filters $w_x$ is applied to the lagged EEG signal $\mathbf{X} \in \mathbb{R}^{T\times L_xC}$, while set of encoder filters $w_s$ is applied to the lagged stimulus envelope $\mathbf{Y}\in \mathbb{R}^{T\times L_s}$, so that the resulting projections are maximally correlated:

\begin{equation}
    \max_{w_x, w_s} \rho(\mathbf{X}w_x, \mathbf{Y}w_y)
    \label{eq:cca}
\end{equation}

The goal of CCA is to maximize the correlation between EEG and stimulus representations and to minimize correlation between components in the latent space where both signals are jointly projected. In this setup, the audio and neural signals are shifted in opposite temporal directions applying a set of time lags. The total number of lags follows the configuration of the original study \cite{decheveigne_decoding_2018}, with $L_x=16$ lags applied to the EEG signal with right zero-padding and $L_s=80$ lags applied with left zero-padding to the stimulus envelope. 

To obtain the canonical coefficients $w_x$ and $w_s$ the CCA matrix $\mathbf{M}$ is first computed as:

\begin{equation}
    \mathbf{M} = \mathbf{C_{xx}^{-\frac{1}{2}}C_{xs}^{}C_{ss}^{-\frac{1}{2}}}
    \label{eq:cca_matrix}
\end{equation}

where $\mathbf{C_{xx}}$ and $\mathbf{C_{ss}}$ denote the EEG and stimulus covariance matrices, respectively, and $\mathbf{C_{xs}}$ denotes the cross covariance matrix. Unlike Ridge Regression, the covariance matrices are normalized. 

The canonical coefficients are then extracted by performing Singular Value Decomposition (SVD) on $\mathbf{M}$ and selecting the first $J$ vectors from the resulting singular matrices $\mathbf{U}$ and $\mathbf{V}$. Afterwards, the whitening matrices, $\mathbf{C_{xx}^{-\frac{1}{2}}}$ and $\mathbf{C_{ss}^{-\frac{1}{2}}}$, are applied to the vectors, yielding the canonical coefficients $w_x^{(j)} \in \mathbb{R}^{L_xC}$ and $w_s^{(j)} \in \mathbb{R}^{L_s}$.

As a result, both EEG and stimuli signals are projected into a $J$ dimensional space. Thus, a multivariate classification method is then required to determine the attended stimulus. Following similar studies \cite{decheveigne_decoding_2018, geirnaert_electroencephalographybased_2021, nguyen_aadnet_2025}, a Linear Discriminant Analysis (LDA) is employed to classify the resulting correlation factors. The optimal $J$ number of components is selected from a range of possible $J$ values that extends up to $min(L_x, L_s)$.

\subsubsection{Auditory Attention Decoding Network (AADNet)}

AADNet proposes deep learning based network that directly identifies the attended stimulus \cite{nguyen_aadnet_2025}. Unlike other conventional AAD methods, AADNet does not perform a two-step procedure in which the correlation is first optimized, followed by classification. Instead, the network takes as input the two isolated audio envelopes present in the auditory scene, together with the corresponding EEG signal, and directly outputs the classification results indicating which envelope was the attended one. As in CCA \cite{decheveigne_decoding_2018}, the signals are projected into a set of temporal sequences, and their correlations are computed to perform classification. Nonetheless, AADNet optimizes the whole network based on the final decoding accuracy using the cross-entropy loss and also considers the unattended source during training. Its network design was inspired by the Inception Network \cite{szegedy_going_2015} and was adapted to the one-dimensional characteristics of AAD data.


The official Github repository \cite{aadnet_2025_github} was employed to define the AADNet model code in this study, ensuring consistency with the author's implementation and training configuration.

\subsection{Causal-Anticausal Temporal Convolutional Network}
\label{sec:ca-tcn model description}

\begin{figure*}[t]
    \centering
        \includegraphics[width=\textwidth]{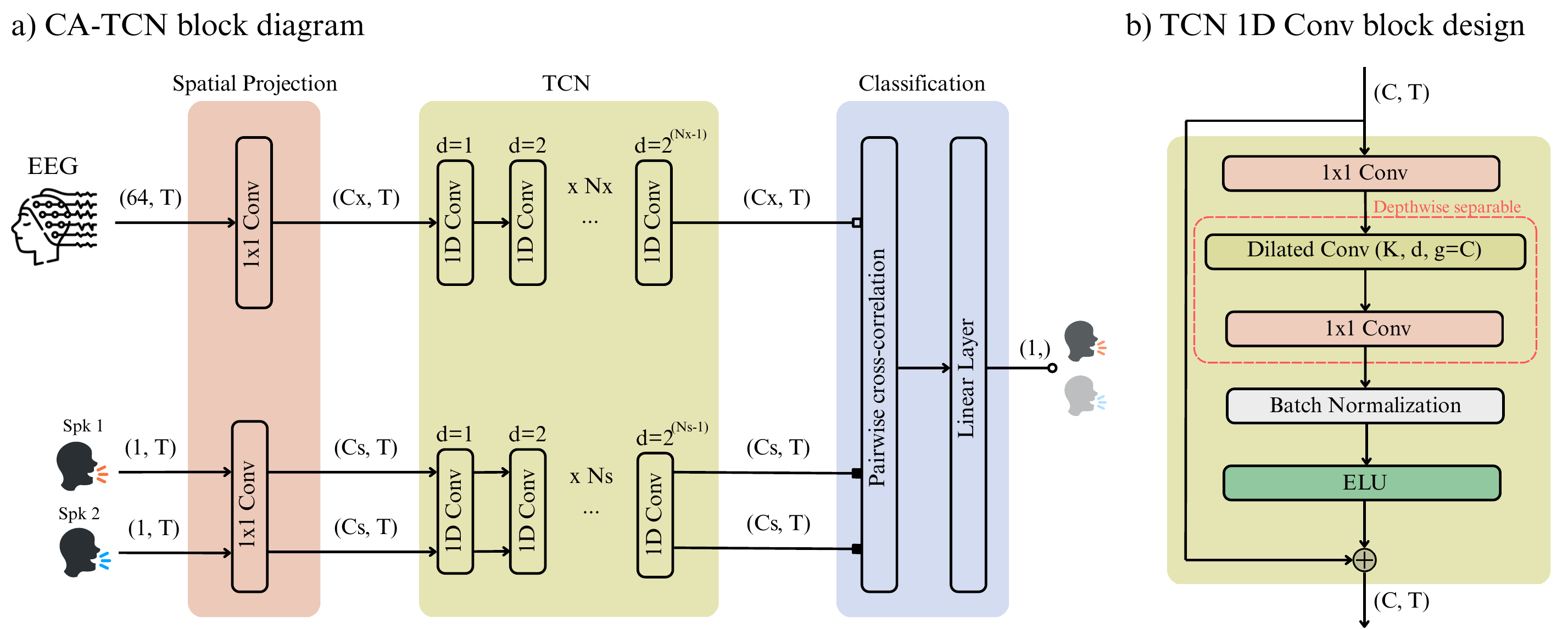}
    \caption{(a) CA-TCN block diagram consisting of a Spatial Projection layer, a Temporal Convolutional Network (TCN) module, and a Classification module. EEG and audio signals are processed separately through two independent branches until classification module. (b) Design of the one-dimensional convolutional block within TCN.}
    \label{fig:Figure_1:TCN_architecture}
\end{figure*}

Aiming to advance the state of entrainment-based AAD models, we introduce CA-TCN, a Causal–Anticausal Temporal Convolutional Network designed to identify the attended speech source in a two-speaker competing scenario. The model processes EEG signals and stimuli separately, resulting in a two-branch architecture: one branch dedicated to EEG and the other to audio, with their outputs finally fused through correlation. The overall architecture can be decomposed into three main stages (see Fig. \ref{fig:Figure_1:TCN_architecture}(a)): (i) a Spatial Projection layer as the stem block, (ii) a Temporal Convolutional Network (TCN) module, (iii) a final classification module.

\subsubsection{Spatial Projection Layer}

The initial processing stage consists of a Spatial Projection layer that maps the inputs into feature spaces with widths \(C_x\) for the EEG branch and \(C_s\) for the stimulus branch. This projection is implemented using \(1 \times 1\) convolutions, which operate exclusively across the channel dimension. For the EEG branch, this operation recombines the input EEG channels, effectively emphasizing the most informative spatial filters. 
Otherwise for the audio branch, this projection expands the original single-channel representation into a higher-dimensional feature space.

\subsubsection{Temporal Convolutional Network (TCN)}

The core processing component of the proposed architecture is based on a Temporal Convolutional Network (TCN). This convolutional architecture has demonstrated strong effectiveness in both audio and EEG processing tasks, including speech separation \cite{luo_convtasnet_2019}, polyphonic music modeling \cite{bai_empirical_2018a}, and EEG classification \cite{ingolfsson_eegtcnet_2020}. It relies on stacks of convolutional layers with exponentially increasing dilation factors ($d=2^{N-1}$), which allow the network to capture long-range temporal dependencies \cite{bai_empirical_2018a}. 

Notably, the receptive field (RF) of a TCN determines the temporal extent over which the network bases its predictions. In our case, its value specifically depends on the kernel size ($K$) and on the number of stacked layers ($N$). This configuration provides a flexible and controllable temporal memory, which can be analytically computed as:
\begin{equation}
    RF = 1 + (K-1)(2^N-1)
    \label{eq:receptive_field}
\end{equation}

For example, a TCN with $K=3$ and $N=4$ yields a RF of 31 samples, corresponding to 484 ms ($f_s=64$ Hz). Importantly, in our AAD context, the RF determines the temporal extent of the EEG and audio processed signals accessible to the model.

The TCN module is composed of multiple stacked convolutional blocks. Each block comprises an initial spatial convolution followed by a depthwise separable convolution, as illustrated in Fig.~\ref{fig:Figure_1:TCN_architecture}(b). The depthwise separable convolution operates separately on the temporal (Dilated conv.) and channel (\(1 \times 1\) conv.) domains, yielding a computationally efficient solution \cite{chollet_xception_2017}. Within each block, Batch Normalization and an ELU activation function are applied prior to the residual addition with the block input, forming a residual connection that stabilizes the training of deep architectures \cite{he_deep_2016}.

\begin{figure*}[t]
    \centering
        \includegraphics[width=\textwidth]{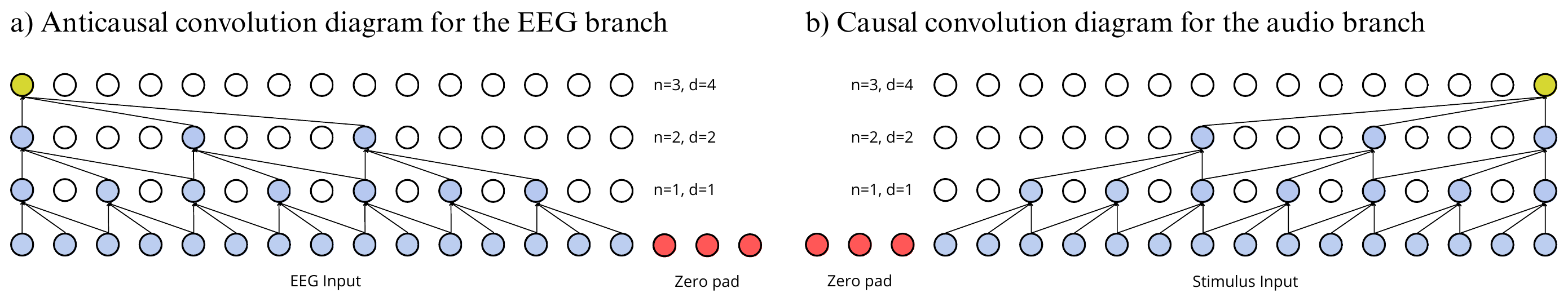}
    \caption{Illustration of dilated convolutions in a three-layer TCN ($N=3$) with exponential dilation factors $d=[1,2,4]$. The yellow sample (decision point) depends on the blue samples conforming its receptive field. Convolutions are applied causally for the stimulus (b) and anticausally for the EEG (a), with zero-padding on the left or right, respectively.}
    \label{fig:TCN_causality}
\end{figure*}

Causality, which defines the temporal direction convolutions are applied, is implemented within the dilated temporal convolutional operation as depicted in Fig.~\ref{fig:TCN_causality}. From our view, it enables a more accurate temporal alignment between stimulus and neural representations accounting for the delay between them. Accordingly, the audio branch employs causal convolutions, ensuring that only past stimulus samples contribute to the representation, and the EEG branch employs anticausal convolutions, incorporating future neural samples relative to the decision point.

Similarly to how we differentiated between causal and anticausal convolutions, the independence between the EEG and audio branch allows us to assign different RFs to each branch. For instance, assuming causality and $K = 3$, five convolutional layers for the EEG branch ($N_x = 5$) and three stacked layers for the stimulus branch ($N_s = 3$) would yield an asymmetric RF of approximately 984~ms over past stimulus samples and 234~ms over future EEG samples according to \ref{eq:receptive_field}.

\subsubsection{Correlation and Classification}

Finally, the resulting temporal sequences extracted by the TCN module are cross-correlated and classified to predict the attended envelope. Specifically, the temporal sequences derived from the EEG branch are cross-correlated with each temporal sequence obtained from the two stimulus candidates. The resulting correlation coefficients are concatenated and fed into a linear classification layer, which outputs a single logit indicating the attended speaker. The overall binary classification task can be interpreted as predicting the attended stimulus index at the network input.

\subsection{Data}
\label{subsec:Data}

Three different datasets were considered in this study. The Jaulab Dataset, which was collected by us, was used for the CA-TCN hyperparameter tuning. KULeuven \cite{das_auditory_2019} and DTU\cite{fuglsang_eeg_2018} datasets are two well-known sets in the AAD literature. Together with the Jaulab Dataset, they were considered for evaluating and comparing the different AAD models.

\subsubsection{Jaulab dataset}
\label{subsec:Jaulab_dataset}

The Jaulab dataset, collected in the Public University of Navarre, comprises EEG recordings from 20 healthy participants while they attended to various news excerpts. The experimental procedure complied with the Declaration of Helsinki and all relevant local regulations. Approval of experimental procedures and protocols was granted by
the Ethics, Animal Experimentation and Biosafety Committee of the Public University of Navarra with number PI-029-20\_CV. EEG signals were recorded using a 64 channel Brain Vision system with a sample frequency of 1000 Hz. The Unified Optimized Layout (UOL) was employed, in which three channels were used to record external signals that were ultimately not included in the analyses. The analysis included a total of 92 trials for each subject. Each trial lasted 26 seconds, and the corresponding audio was presented in different positions via a 24-loudspeaker array arranged in a 2.9-meter-diameter circle within a semi-anechoic chamber. Audio stimuli consists of 192 AI-generated news items, each narrated by one of eight AI-generated voices (four male and four female) produced with Microsoft Azure’s Text-to-Speech tool \cite{azure_2023_framework}. Furthermore, this stimuli were reproduced both from static positions and as moving sources, organized into six distinct presentation blocks that were randomly displayed. In each trial, two news items, narrated with a male and a female voice, were presented simultaneously. Participants were instructed to attend to one audio stream while ignoring the other.

\subsubsection{DTU dataset}

The dataset collected by \textit{Fulsang et al.} at the Technical University of Denmark referred to as the DTU dataset \cite{fuglsang_eeg_2018}, comprises electroencephalography (EEG) recordings from 18 normal-hearing participants who listened to two simultaneous speech streams under varying acoustic conditions. Each trial consisted of 50 seconds of continuous speech presented in acoustic environments with different levels of reverberation. Participants were instructed to selectively attend to one of two spatially separated talkers, one male and one female, while ignoring the other. Trials with a single speaker were excluded in this study, resulting in a total of 60 trials per subject. EEG data was recorded using a 64-channel BioSemi ActiveTwo system, sampled at 512 Hz.

\subsubsection{KULeuven dataset}

This dataset was collected by \textit{Das et al.} at KU Leuven \cite{das_auditory_2019}. It contains electroencephalography (EEG) recordings from 16 normal-hearing subjects, obtained in a soundproof, electromagnetically shielded room using a 64-channel BioSemi ActiveTwo system with a sampling rate of 8196 Hz. Audio stimuli consists of four Dutch short stories narrated by different male speakers and were presented binaurally via insert earphones. During each presentation, subjects were instructed to attend to one ear while ignoring the other. After each presentation, subjects answered multiple-choice questions about the attended story to ensure engagement. In total, each subject completed 20 trials, where the first 8 trials lasted around 6 minutes each and the last 12 around are 2 minutes length, corresponding to approximately 72 minutes of EEG data. Due to differences in the durations of the audio and EEG signals, all trials were trimmed to a uniform length. The available EEG data, already included high-pass filtering, down-sampling to 128 Hz, and artifact removal processing.


\begin{table*}[t]
    \centering
    \caption{Brief information about the data employed in the study}
    \vspace{1ex}
    \label{tab:model_comparison}
    \small
    \resizebox{\textwidth}{!}{
        \begin{tabular*}{\textwidth}{@{\extracolsep{\fill}}cccccc}
            \hline
            & & & \\[-0.8ex]
            \textbf{Dataset} & \textbf{No. of Subjects} & \textbf{No. of Trials} & \textbf{Trial Length} & \textbf{No. of Electrodes} & \textbf{Experimental Conditions} \\[0.8ex]
            \hline
            & & & \\[-0.8ex]
            Jaulab & 20 & 92 & 26 s & 61 &
            \makecell[c]{Loudspeakers, $360^\circ$ \\ Fixed and moving auditory scenarios} \\[0.8ex]
            \hline
            & & & \\[-0.8ex]
            DTU & 18 & 60 & 50 s & 64 &
            \makecell[c]{Insert earphones, $\pm 60^\circ$ \\ 3 different acoustic conditions} \\[0.8ex]
            \hline
            & & & \\[-0.8ex]
            KULeuven & 16 & 20 &
            \makecell[c]{6 min. first 8 trials \\ 2 min. last 12 trials} &
            64 &
            \makecell[c]{Insert earphones, $\pm 90^\circ$ \\ Head-Related Transfer Function \\ and Dichotic audio} \\[0.8ex]
            \hline
            & & & \\[-0.8ex]
        \end{tabular*}
    }
    \vspace{1ex}
\end{table*}

\subsection{Preprocessing}

Data preprocessing followed the same procedure as in \cite{nguyen_aadnet_2025} to ensure comparable results. 

EEG signals were first average-referenced and then band-pass filtered between 0.5 Hz and 32 Hz using a zero-phase finite impulse response (FIR) filter. Subsequently, the data were downsampled to 64 Hz and standardized (z-scored) across all samples and channels within each trial. 

For audio envelope extraction, a Gammatone filterbank was employed. The filterbank was designed following an equivalent rectangular bandwidth (ERB) scale, with center frequencies ranging from 150 Hz to 4 kHz. The resulting sub-band envelopes were compressed using a power-law exponent of 0.6, and then summed to obtain a single broadband envelope. This envelope was subsequently band-pass filtered (0.5–32 Hz), downsampled to 64 Hz, and z-scored.

All filtering and resampling operations were implemented using MNE-Python library \cite{gramfort_meg_2013}. SciPy package \cite{virtanen_scipy_2020} was used for Gammatone filterbank implementation and standardization.

\subsection{Evaluation metrics}

\subsubsection{Classification accuracy}

The main goal of the model is to achieve the highest possible accuracy in determining which speaker is attended. Different window lengths were considered as accuracy varies depending on the number of samples used for the decision. More precisely, we considered six decision window lengths: 1, 2, 5, 10, 25, and 50 seconds. A high accuracy over short decision windows is desirable, as it enables rapid adaptation to shifts in the listener’s attentional focus.

\subsubsection{MESD}

The Minimum Expected Switch Duration (MESD) \cite{geirnaert_interpretable_2020} is a single, interpretable performance metric for AAD that accounts for all obtained accuracies across different decision windows. This metric estimates the time required to perform a gain switch in a hypothetical device following an attentional shift of the user. It is measured in seconds. Lower MESD values indicate a faster switch and better performance. This metric was specifically designed for AAD–gain systems and serves as a unified measure for comparing models, providing a single performance value. The MESD-Python toolbox \cite{mesdtoolbox_2023_github} was used to compute the metric.

\subsubsection{Practical considerations}

Beyond accuracy, some practical considerations were also taken into account to ensure an efficient implementation of the CA-TCN model. The model size and computational efficiency were assessed by counting the number of parameters and the Multiply-Accumulate operations of the model (MACs). We also considered real-time constraints in online processing, which may depend not only on the decision window length but also on the number of samples that constitute the anticausal RF. In fact, this anticausal RF would introduce direct delay on the latency for online operation. Therefore, the larger the anticausal RF, the more constrained would be the model in real-time scenarios. On the other hand, the inference time of the model may be negligible compared to the decision window length.

\subsection{Validation procedure}\label{sec:validation_procedure}

For validating our models, we followed the validation procedure implemented in \cite{nguyen_aadnet_2025}. This procedure follows a cross-trial scheme to ensure there is no within-trial information leakage between sets, preventing the model from learning trial-specific fingerprints \cite{puffay_relating_2023, rotaru_what_2024}. Accordingly, an 8-fold trial-based validation scheme was implemented, dividing the dataset into eight independent trial sets for both subject-specific (SS) and subject-independent (SI) models.

SS models consider data from the evaluated subject during training. In a real use case, SS training would require a data collection process from the specific user. Conversely, SI models implement a leave-one-subject-out (LOSO) validation, in which no information from the evaluated subject is included during training. This approach reproduce a ready-to-use device scenario, capturing all its associated benefits. However, under the SI validation scheme, different trials may contain identical auditory stimuli, potentially leading to audio leakage between the training and test sets. To avoid it, a trial-level inspection was performed, where any training trials sharing the same attended audio track with trials in the test set were excluded from the training data.

For linear models, the same 8-fold trial-based validation scheme described above is followed. As discussed in Section \ref{sec:aad_algorithms_review}, linear approaches require the selection of specific hyperparameters, including the regularization parameter $\lambda$ in Ridge regression and the number of components $J$ in CCA. To this end, a 5-fold inner trial-based validation was performed to determine the optimal values of these hyperparameters.

\subsection{Training procedure}

Training deep learning models typically requires large amounts of data, as a substantially higher number of parameters must be optimized in comparison to simpler linear approaches. For this reason, both AADNet and CA-TCN were first pre-trained using SI data and subsequently fine-tuned using SS data to ensure stable and consistent performance under the SS validation setting. 

Models were trained for 200 epochs using binary cross-entropy loss, with early stopping applied if validation loss did not improve after five consecutive epochs. Five-second decision windows with an overlap of 75\% were used for training. Adam optimizer was employed with a weight decay of $1 \times 10^{-4}$. The learning rate was set to $5 \times 10^{-5}$ for the SI models and reduced by half during fine-tuning for the SS models. Data consisted of neural signals, two candidate envelopes, and a label indicating which of the two envelopes was attended. To prevent bias due to the order of stimulus presentation, the dataset was duplicated and the labels and envelope positions were inverted.

For AADNet, the model was directly obtained from the official Github repository\cite{aadnet_2025_github} and trained following its original configuration. Non-linear models were implemented using the \texttt{PyTorch} framework \cite{paszke_pytorch_2019a}. Conversely, linear models, including Ridge and CCA, were implemented from scratch in \texttt{NumPy} \cite{harris_array_2020} to enable a more suited implementation. In fact, covariance or correlation matrices for each subject and fold were precomputed and used within the respective algorithm.

\subsection{Stability of spatial filters across datasets}
\label{sec:clustering_procedure}

To analyze the stability of the spatial filters learned by the initial spatial projection layer across different dataset, a clustering procedure was employed to group these filters across subjects and folds. We followed the clustering procedure proposed in \cite{mahjoory_convolutional_2024}. First, the spatial filters of each subject were grouped independently into four clusters with k-means. Subsequently, a cross-subject hierarchical clustering was performed on the resulting subject-specific clusters, yielding three global clusters. The cosine angle was used as the similarity metric for cluster linkage. Only the first cross-subject cluster of each dataset was considered as it contained the most informative and numerous spatial distribution. 

To assess the stability of the spatial weighting distributions across different conditions, the cosine similarity was employed to compare the first cluster of each dataset. For the comparison between the Jaulab dataset (61 electrodes) and the other two datasets (64 electrodes), the three missing channels were interpolated using the \texttt{interpolate\_bads()} function from the \texttt{mne} library \cite{gramfort_meg_2013}.

\section{Results}

\subsection{CA-TCN Ablation Study}
\label{subsec:ablation_study_results}

\begin{table*}
\centering
\caption{Ablation study results obtained by progressively incorporating architectural components, starting from a correlation-and-classification-only baseline and culminating in the final optimized CA-TCN model. The accuracy corresponds with the validation accuracy obtained using a five second decision window for the subject-independent setting on the Jaulab dataset.}
\label{tab:ablation}
\small
\begin{tabular}{@{\extracolsep{\fill}}lcccccccc}
\toprule
& & & \\[-0.8ex]
\textbf{Model State} & \textbf{Cx} & \textbf{Cs} & \textbf{Nx} & \textbf{Ns} & \textbf{K} & \textbf{Parameters} & \textbf{MACs} & \textbf{Accuracy (5s)} \\[0.8ex]
\midrule
& & & \\[-0.8ex]
Raw Input          & \ding{53} & \ding{53} & \ding{53} & \ding{53} & \ding{53} & 0.1K & 0.1K & 53.9\% \\[0.8ex]
+ Spatial Projection & 32 & 32 & \ding{53} & \ding{53} & \ding{53} & 4.1K & 2.1M & 55.4\%\\[0.8ex]
+ TCN               & 32 & 32 & 5  & 5 & 3 & 27.1K & 37.6M & 71.8\%\\[0.8ex]
+ Causality         & 32 & 32 & 4  & 4 & 3 & 22.5K & 30.5M & 72.0\%\\[0.8ex]
+ Asymmetric RF (CA-TCN)    & 32 & 32 & 3 & 5 & 3 & 22.5K & 32.9M & 72.1\%\\[0.8ex]
\bottomrule
\end{tabular}
\end{table*}

As introduced in Section~\ref{sec:ca-tcn model description}, the CA-TCN architecture is composed of several modules incorporating distinct design choices. An ablation study was conducted to evaluate the contribution of each stage and architectural component to the overall decoding performance. Starting from a classifier-only baseline, network components were progressively incorporated until reaching the final optimized CA-TCN model. This analysis was performed by monitoring the validation loss and decoding accuracy under the SI setting on the Jaulab dataset, using a 5 s decision window.

The results reported in Table~\ref{tab:ablation} summarize the marginal contribution of each architectural component. The first row corresponds to a classifier-only configuration, in which the raw input signals are fed directly into the correlation and classification module. As expected, this baseline operates close to chance level. When the spatial projection layer is included in isolation, performance remains near chance, although a slight improvement of approximately 1.5 percentage points in decoding accuracy is observed. 

Consistent with our expectations, the largest contribution to the model’s performance corresponded with the inclusion of the TCN module, which yielded a improvement of 16.4\% in decoding accuracy. With the incorporation of causal convolutions and the use of an asymmetric RF for the EEG and stimulus branches, modest performance gains are observed, with increases of 0.2\% and 0.1\% respectively. However, with the incorporation of causality, the model size was reduced by 4.6K parameters, and the number of MACs decreased by 7.1M, corresponding to approximately 17\% and 19\%, respectively. Moreover, the asymmetry between RFs reduced the number of samples considered by the anticausal EEG branch from 31 to 15, corresponding to 484 ms and 234 ms respectively.

\subsection{AAD model performance comparison}

In this section, CA-TCN is evaluated together with the other AAD methods presented in Section \ref{sec:aad_algorithms_review} across the three datasets described in \ref{subsec:Data}.

\label{subsec:comparison_performance_results}

\subsubsection{Subject-Independent (SI) models}

\begin{table*}
    \centering
    \caption{Model performance comparison for Subject-Independent validation across three datasets}
    \vspace{1ex}
    \label{tab:multi_dataset_si_comparison}
    \resizebox{\textwidth}{!}{
        \begin{tabular}{ccccccccc}
            \toprule
            \multirow{2}{*}{Dataset} & \multirow{2}{*}{Model} & \multicolumn{6}{c}{Accuracy (\%)} & \multirow{2}{*}{MESD (s)} \\
            \cmidrule(lr){3-8}
            & & 1s & 2s & 5s & 10s & 25s & 50s & \\
            \midrule
            \multirow{4}{*}{Jaulab} & Ridge   & 54.2 $\pm$ 1.9 & 56.0 $\pm$ 2.7 & 59.6 $\pm$ 3.9 & 63.5 $\pm$ 5.4 & 69.1 $\pm$ 7.4  & 74.6 $\pm$ 9.8  & $> 100$ \\
            & CCA     & 55.1 $\pm$ 1.9 & 58.3 $\pm$ 3.2 & 63.3 $\pm$ 5.0 & 68.3 $\pm$ 7.1 & 75.4 $\pm$ 9.6  & 81.4 $\pm$ 11.5 & $> 100$ \\
            & AADNet & 57.5 $\pm$ 2.6 & 61.6 $\pm$ 3.6 & 68.6 $\pm$ 5.1 & 74.5 $\pm$ 6.5 & 82.3 $\pm$ 7.7  & 86.4 $\pm$ 8.6  & 26.9 $\pm$ 16.7 \\
            & CA-TCN  & \textbf{58.0} $\pm$ 2.4 & \textbf{62.5} $\pm$ 4.0 & \textbf{69.6} $\pm$ 5.6 & \textbf{75.4} $\pm$ 7.1 & \textbf{84.0} $\pm$ 8.8  & \textbf{88.5} $\pm$ 7.7  & \textbf{25.1} $\pm$ 20.8 \\
            \midrule
            \multirow{4}{*}{DTU} & Ridge   & 56.0 $\pm$ 2.8 & 60.1 $\pm$ 4.2 & 65.6 $\pm$ 6.4 & 70.2 $\pm$ 7.9 & 78.1 $\pm$ 10.1 & 83.6 $\pm$ 11.2 & 82.4 $\pm$ 129.8 \\
            & CCA     & 56.5 $\pm$ 2.6 & 60.6 $\pm$ 4.4 & 67.0 $\pm$ 6.4 & 72.6 $\pm$ 7.7 & 80.7 $\pm$ 9.6  & 87.7 $\pm$ 10.5 & 53.4 $\pm$ 73.9 \\
            & AADNet & 57.5 $\pm$ 2.4 & 62.1 $\pm$ 4.0 & 69.4 $\pm$ 5.9 & 75.8 $\pm$ 7.2 & 84.3 $\pm$ 8.2  & 90.9 $\pm$ 8.6  & \textbf{32.2} $\pm$ 45.9 \\
            & CA-TCN  & \textbf{58.3} $\pm$ 2.8 & \textbf{63.4} $\pm$ 4.4 & \textbf{70.3} $\pm$ 6.9 & \textbf{76.7} $\pm$ 7.7 & \textbf{85.8} $\pm$ 8.3  & \textbf{92.5} $\pm$ 8.0  & 34.7 $\pm$ 68.4 \\
            \midrule
            \multirow{4}{*}{KULeuven} & Ridge   & 54.4 $\pm$ 1.7 & 57.0 $\pm$ 2.5 & 61.5 $\pm$ 3.8 & 65.3 $\pm$ 5.4 & 71.3 $\pm$ 8.0  & 76.4 $\pm$ 10.5 & $> 100$ \\
            & CCA    & 55.2 $\pm$ 1.5 & 58.3 $\pm$ 2.4 & 63.7 $\pm$ 3.4 & 68.6 $\pm$ 4.9 & 75.1 $\pm$ 7.2  & 81.3 $\pm$ 8.5  & $> 100$ \\
            & AADNet & 56.0 $\pm$ 2.0 & 60.2 $\pm$ 3.0 & 66.3 $\pm$ 4.2 & 71.5 $\pm$ 5.7 & 79.0 $\pm$ 8.4  & 84.1 $\pm$ 8.2  & 34.3 $\pm$ 22.1 \\
            & CA-TCN  & \textbf{56.8} $\pm$ 2.1 & \textbf{61.1} $\pm$ 3.0 & \textbf{67.9} $\pm$ 4.2 & \textbf{73.4} $\pm$ 5.5 & \textbf{81.9} $\pm$ 6.7  & \textbf{87.3} $\pm$ 6.9  & \textbf{27.5} $\pm$ 16.1 \\
            \bottomrule
        \end{tabular}
    }
    \vspace{1ex}
\end{table*}

\begin{figure*}
    \centering
    \includegraphics[width=\textwidth]{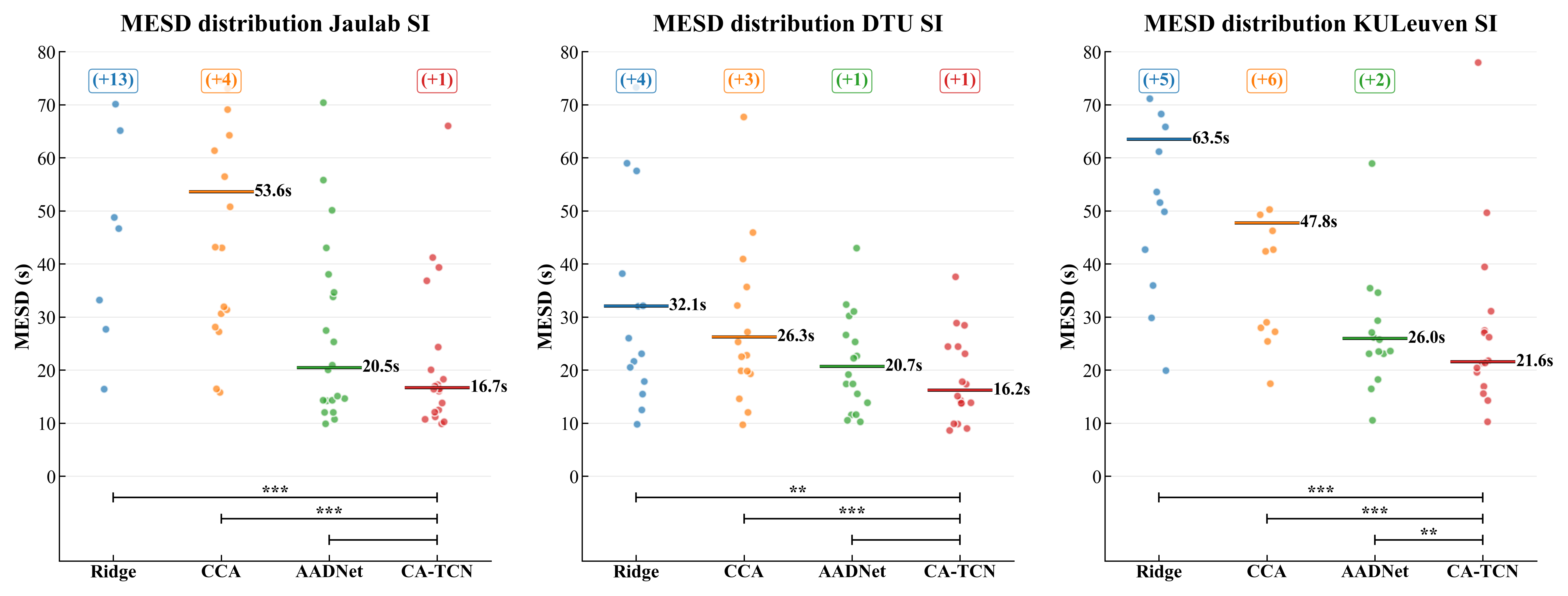}
    \caption{Subject-independent (SI) distribution of MESD across the three datasets considered in this study. Each point corresponds to the MESD value obtained for a given subject, and median values are highlighted in the distributions. The number of subjects whose MESD exceeds 80 s are indicated at the top of the distributions in parentheses. Statistical analysis: ***: $p < 0.001$, **: $0.001\leq p<0.01$, *: $0.01\leq p<0.05$, None: $p\geq0.05$.}
    \label{fig:Figure_3_MESD_SI}
\end{figure*}

Table~\ref{tab:multi_dataset_si_comparison} summarizes the SI results obtained for each AAD model. Across all evaluated decision window lengths and datasets, CA-TCN consistently achieved the highest decoding accuracy among all models. For instance, on the Jaulab dataset, accuracy increased from 58.0\% with a 1 s decision window to 88.5\% with a 50 s window. Similar trends were observed for the KULeuven dataset, where accuracy ranged from 56.8\% to 87.3\%, and for the DTU dataset, where performance improved from 58.3\% to 92.5\%. Moreover, the relative performance gains of CA-TCN over the next best-performing model, AADNet, became more pronounced as the decision window length increased. For example, on the KULeuven dataset, CA-TCN provided a modest improvement of 0.8\% for 1 s windows, whereas this gain increased to 3.2\% for 50 s windows.

The right column of Table \ref{tab:multi_dataset_si_comparison} reports the mean MESD values across folds and subjects. CA-TCN achieved the lowest mean MESD on the Jaulab and KULeuven datasets, with marginal reductions of 1.8 s and 6.8 s, respectively, compared to AADNet. Despite outperforming AADNet across all decision window lengths, CA-TCN did not yield a lower mean MESD on the DTU dataset; instead, AADNet achieved a value that was 2.5 s lower. This behavior may be attributed to the higher inter-subject variability observed on the DTU dataset when using CA-TCN. 

On the othe hand, linear models consistently exhibited poorer performance than deep learning approaches, highlighting the inherent limitations of linear decoding methods for cross-subject AAD.

Fig. ~\ref{fig:Figure_3_MESD_SI} further illustrates the subject distribution of the MESD values for each dataset by averaging across folds. Paired Wilcoxon tests were performed to assess significant differences between MESD values when comparing CA-TCN with other models. To account for multiple comparisons, a Bonferroni correction was applied to adjust the resulting p-values. Statistical comparisons between the distributions of CA-TCN and AADNet revealed a significant difference only for the KULeuven dataset ($p = 0.002$), whereas no significant differences were observed for the Jaulab and DTU datasets.

\subsubsection{Subject-Specific (SS) models}

\begin{table*}
    \centering
    \caption{Model performance comparison for Subject-Specific validation across three datasets}
    \vspace{1ex}
    \label{tab:multi_dataset_ss_comparison}
    \resizebox{\textwidth}{!}{
        \begin{tabular}{ccccccccc}
            \toprule
            \multirow{2}{*}{Dataset} & \multirow{2}{*}{Model} & \multicolumn{6}{c}{Accuracy (\%)} & \multirow{2}{*}{MESD (s)} \\
            \cmidrule(lr){3-8}
            & & 1s & 2s & 5s & 10s & 25s & 50s & \\
            \midrule
            \multirow{4}{*}{Jaulab} & Ridge   & 56.7 $\pm$ 2.7 & 60.7 $\pm$ 4.0 & 67.2 $\pm$ 6.0 & 73.5 $\pm$ 7.5 & 82.2 $\pm$ 8.4 & 89.0 $\pm$ 8.6 & 58.7 $\pm$ 109.4 \\
            & CCA     & 57.5 $\pm$ 2.9 & 62.1 $\pm$ 4.7 & 69.2 $\pm$ 6.9 & 75.5 $\pm$ 8.2 & 85.1 $\pm$ 8.9 & 92.3 $\pm$ 7.9 & 78.1 $\pm$ 205.1 \\
            & AADNet & 59.4 $\pm$ 2.9 & 64.8 $\pm$ 4.8 & 73.0 $\pm$ 6.7 & 79.1 $\pm$ 8.1 & 86.3 $\pm$ 8.0 & 89.7 $\pm$ 7.1 & 20.1 $\pm$ 17.2 \\
            & CA-TCN  & \textbf{60.4} $\pm$ 3.1 & \textbf{66.1} $\pm$ 4.5 & \textbf{74.3} $\pm$ 6.3 & \textbf{81.3} $\pm$ 7.0 & \textbf{89.2} $\pm$ 6.2 & \textbf{91.6} $\pm$ 4.5 & \textbf{15.4} $\pm$ 10.4 \\
            \midrule
            \multirow{4}{*}{DTU} & Ridge   & 58.8 $\pm$ 2.1 & 65.4 $\pm$ 3.2 & 73.3 $\pm$ 4.7 & 79.8 $\pm$ 5.4 & 87.9 $\pm$ 5.2 & 94.0 $\pm$ 4.3 & 16.6 $\pm$ 7.6 \\
            & CCA     & 58.9 $\pm$ 2.2 & 64.8 $\pm$ 3.3 & 72.8 $\pm$ 4.8 & 79.0 $\pm$ 5.5 & 87.9 $\pm$ 5.7 & 93.8 $\pm$ 4.3 & 18.6 $\pm$ 8.7 \\
            & AADNet & 59.6 $\pm$ 2.3 & 65.8 $\pm$ 3.3 & 74.3 $\pm$ 4.9 & 81.2 $\pm$ 5.9 & 90.2 $\pm$ 5.9 & 95.8 $\pm$ 4.1 & 13.9 $\pm$ 4.5 \\
            & CA-TCN  & \textbf{60.5} $\pm$ 2.8 & \textbf{67.3} $\pm$ 4.6 & \textbf{76.6} $\pm$ 6.2 & \textbf{83.4} $\pm$ 5.8 & \textbf{92.4} $\pm$ 4.7 & \textbf{96.6} $\pm$ 3.3 & \textbf{13.2} $\pm$ 6.4 \\
            \midrule
            \multirow{4}{*}{KULeuven} & Ridge   & 57.8 $\pm$ 2.8 & 61.9 $\pm$ 4.2 & 67.9 $\pm$ 6.1 & 73.8 $\pm$ 7.3 & 82.0 $\pm$ 8.0 & 87.4 $\pm$ 8.5 & 41.9 $\pm$ 49.6 \\
            & CCA     & 58.3 $\pm$ 2.5 & 63.0 $\pm$ 4.0 & 70.3 $\pm$ 5.4 & 76.8 $\pm$ 6.4 & 85.2 $\pm$ 7.1 & 91.5 $\pm$ 6.9 & 26.3 $\pm$ 21.0 \\
            & AADNet & 58.1 $\pm$ 2.6 & 63.4 $\pm$ 4.1 & 71.1 $\pm$ 5.2 & 77.2 $\pm$ 6.8 & 86.5 $\pm$ 7.3 & 91.0 $\pm$ 8.1 & 21.7 $\pm$ 16.3 \\
            & CA-TCN  & \textbf{59.1} $\pm$ 3.0 & \textbf{64.4} $\pm$ 4.1 & \textbf{72.3} $\pm$ 5.4 & \textbf{78.9} $\pm$ 6.4 & \textbf{87.5} $\pm$ 7.5 & \textbf{91.8} $\pm$ 7.1 & \textbf{19.3} $\pm$ 16.8 \\
            \bottomrule
        \end{tabular}
    }
    \vspace{1ex}
\end{table*}

\begin{figure*}
    \centering
    \includegraphics[width=\textwidth]{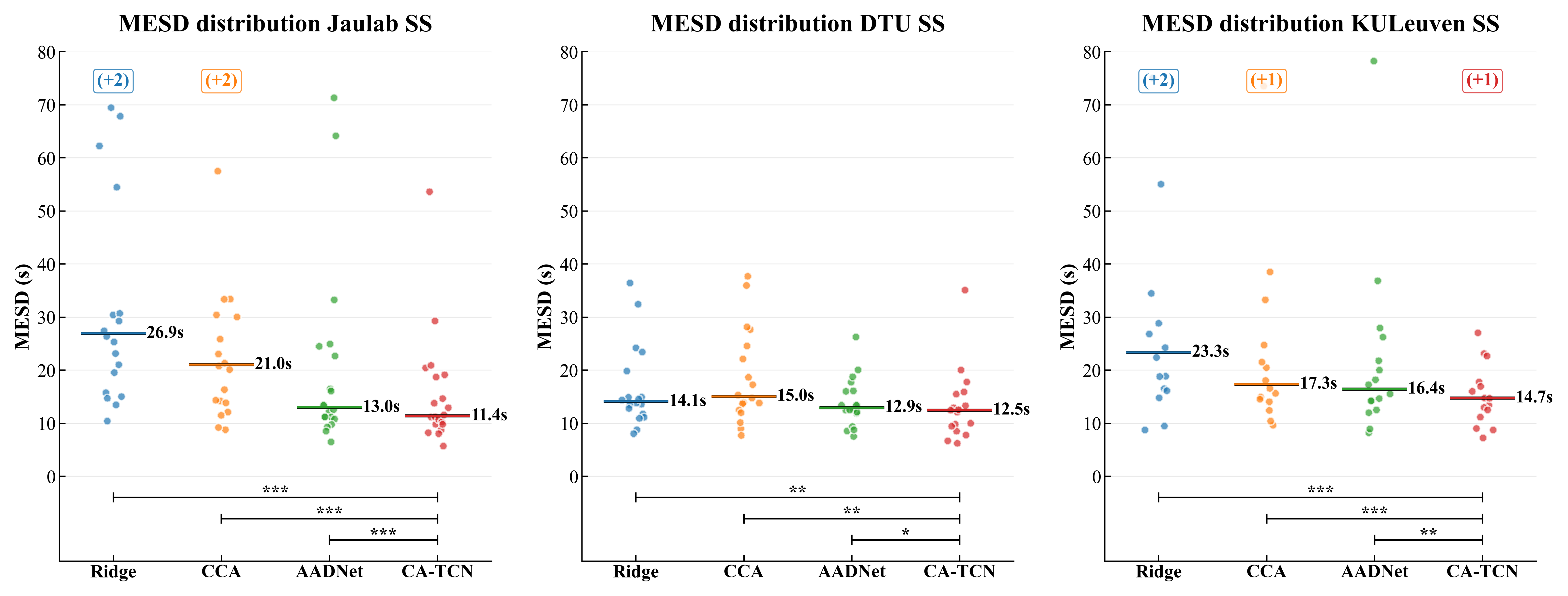}
    \caption{Subject-Specific (SS) distribution of MESD across the three datasets considered in this study. Each point corresponds to the average MESD obtained for a given subject, and median values are highlighted in the distributions.}
    \label{fig:Figure_4_MESD_SS}
\end{figure*}

Table~\ref{tab:multi_dataset_ss_comparison} summarizes the SS results obtained across all datasets considered in the study. As for the SI validation scheme, CA-TCN achieved the highest decoding accuracies across all the evaluated settings. On the Jaulab dataset, CA-TCN accuracy increased from 60.4\% for a 1 s window to 91.6\% for a 50 s window. Similar performance trends were observed for the DTU dataset, with accuracy ranging from 60.5\% to 96.6\%, and for the KULeuven dataset, where performance improved from 59.1\% to 91.8\%.

Beyond accuracy, the mean MESD values reported in Table \ref{tab:multi_dataset_ss_comparison} further highlight the benefits of CA-TCN in the SS setting. Relative to AADNet, CA-TCN achieved consistent reductions in MESD across all evaluated datasets, with improvements of 2.4 s on KULeuven, 0.7 s on DTU, and 4.7 s on Jaulab. Under this validation scheme, linear models attained performance levels comparable to those of deep learning approaches on the KULeuven and DTU datasets; however, their performance on the Jaulab dataset remained substantially inferior. 

Figure~\ref{fig:Figure_4_MESD_SS} illustrates the distribution of MESD values across subjects for each dataset under the SS evaluation. Statistical comparisons between CA-TCN and AADNet revealed significant differences across all datasets when SS data were used during training. The weakest statistical evidence was observed for the DTU dataset ($p = 0.047$), followed by KULeuven ($p = 0.006$), while the strongest evidence was found for the Jaulab dataset ($p = 8 \times 10^{-5}$).

\subsection{Spatial filters}
\label{sec:results_spatial_weighting}

\begin{figure*}[t]
    \centering
    \includegraphics[width=\textwidth]{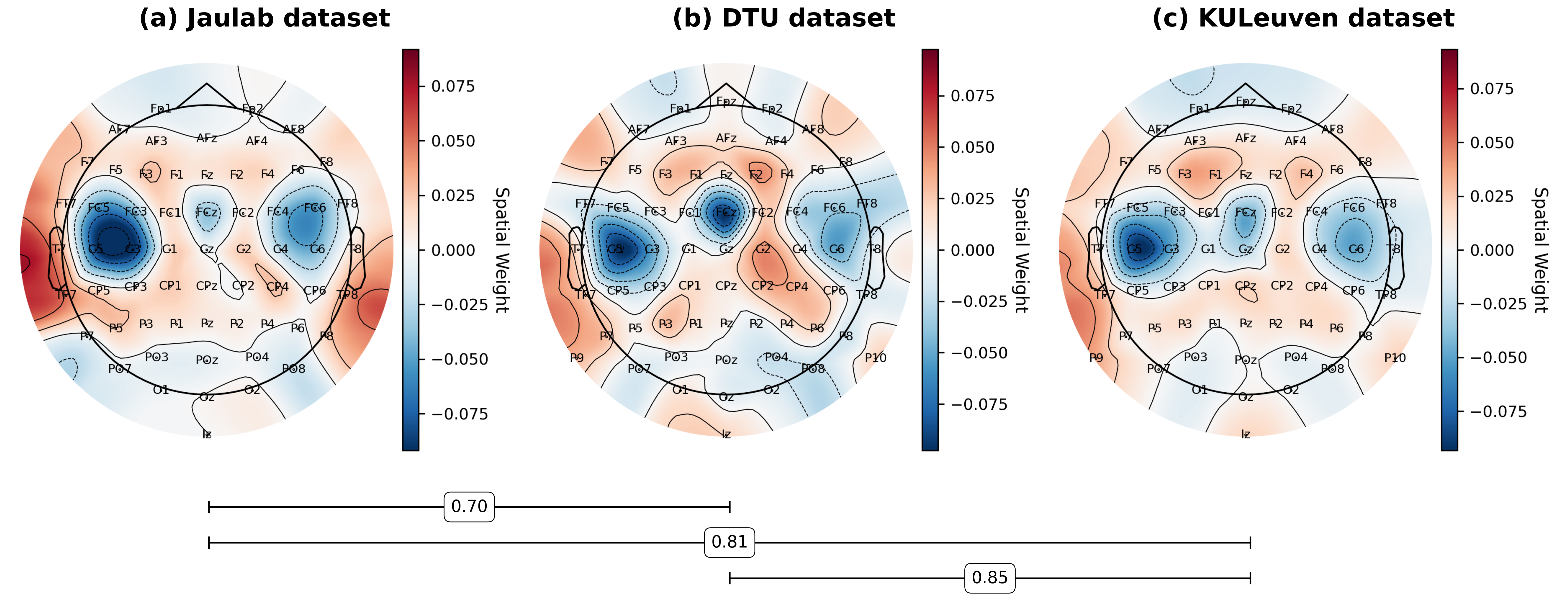}
    \caption{Topographic maps of the first cluster obtained by grouping the Spatial Projection filters from the CA-TCN model across the Jaulab (a), DTU (b), and KULeuven (c) datasets. The values below the topographic maps indicate the cosine similarity between each pair of clusters. These maps reveal a predominance of fronto-central, left-lateralized areas.}
    \label{fig:Figure_topomap_All}
\end{figure*}

The first clusters obtained for each dataset, as described in Section \ref{sec:clustering_procedure}, are illustrated in Fig.~\ref{fig:Figure_topomap_All} as topographic maps. These spatial centroids exhibited a highly consistent pattern across datasets, characterized by a predominant contribution from left-lateralized central regions, peaking at electrode C5. Consistency across datasets was further supported by the high inter-dataset cosine similarities: $0.64$ for Jaulab–DTU, $0.75$ for Jaulab–KULeuven, and $0.90$ for DTU–KULeuven.

\section{Discussion}

\subsection{Unified Temporal Modeling}

CA-TCN consistently improved accuracy compared with the next best-performing model, AADNet. For SI models, the improvement ranged from 0.5\% to 3.2\%, whereas for SS models it ranged from 0.8\% to 2.9\%. These results highlight the benefits of the proposed CA-TCN architecture and its unified temporal processing through a TCN design, in contrast to the multi-branch paradigm adopted by AADNet.

In fact, despite incorporating several established design principles in modern convolutional neural networks, including dilated convolutions, residual connections \cite{he_deep_2016}, and depthwise separable convolutions \cite{chollet2017xception}, a defining characteristic of the proposed CA-TCN architecture is its overall structural simplicity. The network is defined by a compact set of global hyperparameters and relies on the progressive increase of the dilation factor across layers to effectively model temporal dependencies at multiple scales. 

In contrast, AADNet \cite{nguyen_aadnet_2025} adopts a more heterogeneous architectural design based on multiple parallel branches, each explicitly tailored to capture specific temporal characteristics through predefined kernel sizes, channel configurations, or pointwise convolutions. We hypothesize that this architectural fragmentation may limit the model’s ability to jointly exploit temporal information across scales, which could partly explain the lower accuracies observed. Moreover, the unified design of CA-TCN relies on a compact and smaller set of hyperparameters, which simplifies the exploration of different network configurations.

\subsection{Improving efficiency and minimizing latency}
\label{sec:Efficient AAD model}

The incorporation of causality within convolutions, as described in Section~\ref{sec:ca-tcn model description}, not only yielded a slight improvement in decoding performance but also substantially reduced the model complexity. As shown in Table~\ref{tab:ablation}, a model composed of four stacked causal convolutional layers ($N=4$) outperformed the deeper version with an additional layer ($N=5$) that did not enforce causality. These results revealed the effectiveness of integrating causality, showing the benefits of designing a task-aligned model to account for the delay between stimuli and the neural response.

Moreover, the inclusion of asymmetric RFs led to a shorter RF for the EEG branch (234 ms) and a broader past-oriented RF for the stimulus branch (-984 ms). Therefore, compared with its previous version ($N_x=4$ and $N_s=4$), the model reduced its reliance on EEG samples while leveraging an extended audio-envelope context. This configuration is particularly advantageous for real-time attention monitoring, where a reduction in the anticausal RF directly reduces its associated delay. We remark that the final anticausal RF would introduce an extra delay of 234 ms on the latency for online processing, which may end up being very short relative to the decision window used. Indeed, even when compared with the shortest window in the study (1 s), this delay represents less than a quarter of its duration.

\subsection{Spatial stability across different conditions}

The spatial distributions revealing the first cluster across datasets highlight a consistent pattern, with particular emphasis on electrode C5. Notably, the high cosine similarities demonstrate stable spatial filters across datasets, indicating that the learned representations are robust to variations in recording setups and acoustic environments. 

This spatial distribution also aligns with previous AAD studies. For instance, electrodes FC5 and C3 (C5 was not included in that study) were found to be the most effective when added to ear-EEG systems \cite{geirnaert_direct_2025a}. Similarly, when a leave-one-channel-out validation was performed in the AADNet study, electrode C5 was found to be the most consistently informative one \cite{nguyen_aadnet_2025}. Although this information about the spatial filters cannot be unambiguously interpreted in terms of the underlying neurophysiological mechanisms of auditory attention \cite{haufe_interpretation_2014}, this left-lateralized fronto-central dependence, with a peak at electrode C5, lies close to the auditory cortex and the left superior temporal gyrus. These regions are associated with speech-envelope tracking \cite{kubanek_tracking_2013} and with the coupling to the attended stimulus in a Cocktail Party scenario \cite{vanderghinst_left_2016}. 

\subsection{Limitations and future work}

The current results are inherently constrained to controlled, non-realistic AAD conditions, as analyses were conducted post hoc using clean speech envelopes and experimentally controlled stimuli. Within these conditions, CA-TCN consistently outperforms previously proposed models, demonstrating the advantages of the non-linear architecture. However, the model’s performance in more realistic scenarios—incorporating speech separation algorithms \cite{luo_convtasnet_2019, tanveer_deep_2024}, real-time processing \cite{hjortkjaer_realtime_2025, aroudi_closedloop_2021}, naturalistic acoustic environments \cite{straetmans_neural_2024}, or miniaturized EEG acquisition devices \cite{geirnaert_direct_2025a}—remains to be evaluated. Since most real-world AAD studies rely on linear decoding models, both objective and subjective evaluations of integrating a non-linear architecture such as CA-TCN into these pipelines represent a valuable direction for future research. In particular, it would be of special interest to evaluate the model in a real-world SI paradigm, where CA-TCN has demonstrated a substantial performance gap compared to linear models. Determining whether the observed performance gains persist under realistic conditions would be a critical step toward the practical deployment of neuro-steered hearing devices.

\section{Conclusion}

This work proposes CA-TCN, a TCN-based deep neural network designed to directly classify the attended speaker in a two-speaker competing scenario. The core component of the network is based on a TCN, which relies on a stack of convolutional layers with an exponentially increasing dilation factor. Importantly, the network differs in the way it processes EEG and audio signals. Neural signals are processed anticausally, basing its predictions on future samples within a limited RF, whereas stimulus signals are processed causally, relying only on past samples within an extended RF. The ablation study revealed the importance of the TCN module and also considered the incorporation of causality and asymmetry between RFs. These design choices helped improve the efficiency and aimed to minimize the online processing latency. Across all the considered settings, CA-TCN consistently outperformed other established AAD methods and significantly improved the MESD distributions in four of the six evaluated settings compared to the next best-performing model. Furthermore, the spatial filters from the initial EEG convolutional layer revealed stable distributions across datasets, highlighting the model’s spatial robustness to different experimental conditions. Nevertheless, this study still relied on controlled experimental data, and further analysis involving more realistic scenarios is needed to reveal the true potential of the model design.

\section*{References}

\begin{list}{[\arabic{enumi}]}{\settowidth\labelwidth{[99]}\leftmargin\labelwidth
\advance\leftmargin\labelsep
\usecounter{enumi}}
\small

\bibitem{cherry_experiments_1953}
E.~C. Cherry, ``Some {{Experiments}} on the {{Recognition}} of {{Speech}}, with {{One}} and with {{Two Ears}},'' \emph{The Journal of the Acoustical Society of America}, vol.~25, no.~5, pp. 975--79, Sep. 1953.

\bibitem{lesica_why_2018}
N.~A. Lesica, ``Why {{Do Hearing Aids Fail}} to {{Restore Normal Auditory Perception}}?'' \emph{Trends in Neurosciences}, vol.~41, no.~4, pp. 174--85, Apr. 2018.

\bibitem{luo_convtasnet_2019}
Y.~Luo and N.~Mesgarani, ``Conv-{{TasNet}}: {{Surpassing Ideal Time-Frequency Magnitude Masking}} for {{Speech Separation}},'' \emph{IEEE/ACM Trans. Audio Speech Lang. Process.}, vol.~27, no.~8, pp. 1256--66, Aug. 2019.

\bibitem{wang_deep_2017}
D.~Wang, ``Deep learning reinvents the hearing aid,'' \emph{IEEE Spectrum}, vol.~54, no.~3, pp. 32--7, Mar. 2017.

\bibitem{hjortkjaer_realtime_2025}
J.~Hjortkj{\ae}r, D.~D.~E. Wong, A.~Catania, J.~{M{\"a}rcher-R{\o}rsted}, E.~Ceolini, S.~A. Fuglsang, I.~Kiselev, G.~Di~Liberto, S.-C. Liu, T.~Dau, M.~Slaney, and A.~{de Cheveign{\'e}}, ``Real-time control of a hearing instrument with {{EEG-based}} attention decoding,'' \emph{J. Neural Eng.}, vol.~22, no.~1, Feb. 2025.

\bibitem{geirnaert_electroencephalographybased_2021}
S.~Geirnaert, S.~Vandecappelle, E.~Alickovic, A.~{de Cheveigne}, E.~Lalor, B.~T. Meyer, S.~Miran, T.~Francart, and A.~Bertrand, ``Electroencephalography-{{Based Auditory Attention Decoding}}: {{Toward Neurosteered Hearing Devices}},'' \emph{IEEE Signal Processing Magazine}, vol.~38, no.~4, pp. 89--102, Jul. 2021.

\bibitem{aiken_human_2008}
S.~J. Aiken and T.~W. Picton, ``Human {{Cortical Responses}} to the {{Speech Envelope}},'' \emph{Ear \& Hearing}, vol.~29, no.~2, pp. 139--57, Apr. 2008.

\bibitem{osullivan_attentional_2015}
J.~A. O'Sullivan, A.~J. Power, N.~Mesgarani, S.~Rajaram, J.~J. Foxe, B.~G. {Shinn-Cunningham}, M.~Slaney, S.~A. Shamma, and E.~C. Lalor, ``Attentional {{Selection}} in a {{Cocktail Party Environment Can Be Decoded}} from {{Single-Trial EEG}},'' \emph{Cerebral Cortex}, vol.~25, no.~7, pp. 1697--1706, Jul. 2015.

\bibitem{akram_robust_2016}
S.~Akram, A.~Presacco, J.~Z. Simon, S.~A. Shamma, and B.~Babadi, ``Robust decoding of selective auditory attention from {{MEG}} in a competing-speaker environment via state-space modeling,'' \emph{NeuroImage}, vol. 124, pp. 906--17, Jan. 2016.

\bibitem{dijkstra_identifying_2015}
K.~Dijkstra, P.~Brunner, A.~Gunduz, W.~Coon, A.~Ritaccio, J.~Farquhar, and G.~Schalk, ``Identifying the attended speaker using electrocorticographic ({{ECoG}}) signals,'' \emph{Brain-Computer Interfaces}, vol.~2, no.~4, pp. 161--73, Oct. 2015.

\bibitem{raghavan_improving_2024}
V.~S. Raghavan, J.~O'Sullivan, J.~Herrero, S.~Bickel, A.~D. Mehta, and N.~Mesgarani, ``Improving auditory attention decoding by classifying intracranial responses to glimpsed and masked acoustic events,'' \emph{Imaging Neurosci.}, vol.~2, 2024.

\bibitem{meyer_neural_2018}
L.~Meyer, ``The neural oscillations of speech processing and language comprehension: State of the art and emerging mechanisms,'' \emph{European Journal of Neuroscience}, vol.~48, no.~7, pp. 2609--21, 2018.

\bibitem{nguyen_aadnet_2025}
N.~D.~T. Nguyen, H.~Phan, S.~Geirnaert, K.~Mikkelsen, and P.~Kidmose, ``{{AADNet}}: {{An End-to-End Deep Learning Model}} for {{Auditory Attention Decoding}},'' \emph{IEEE Transactions on Neural Systems and Rehabilitation Engineering}, vol.~33, pp. 2695--2706, 2025.

\bibitem{aroudi_closedloop_2021}
A.~Aroudi, E.~Fischer, M.~Serman, H.~Puder, and S.~Doclo, ``Closed-{{Loop Cognitive-Driven Gain Control}} of {{Competing Sounds Using Auditory Attention Decoding}},'' vol.~14, no.~10, p. 287.

\bibitem{meyer_synchronous_2020}
L.~Meyer, Y.~Sun, and A.~E. Martin, ``Synchronous, but not entrained: Exogenous and endogenous cortical rhythms of speech and language processing,'' \emph{Language, Cognition and Neuroscience}, vol.~35, no.~9, pp. 1089--99, Nov. 2020.

\bibitem{viswanathan_electroencephalographic_2019}
V.~Viswanathan, H.~M. Bharadwaj, and B.~G. {Shinn-Cunningham}, ``Electroencephalographic {{Signatures}} of the {{Neural Representation}} of {{Speech}} during {{Selective Attention}},'' \emph{eNeuro}, vol.~6, no.~5, Oct. 2019.

\bibitem{wong_comparison_2018}
D.~D.~E. Wong, S.~A. Fuglsang, J.~Hjortkj{\ae}r, E.~Ceolini, M.~Slaney, and A.~De~Cheveign{\'e}, ``A {{Comparison}} of {{Regularization Methods}} in {{Forward}} and {{Backward Models}} for {{Auditory Attention Decoding}},'' \emph{Front. Neurosci.}, vol.~12, p. 531, Aug. 2018.

\bibitem{decheveigne_decoding_2018}
A.~{de Cheveign{\'e}}, D.~D.~E. Wong, G.~M. Di~Liberto, J.~Hjortkj{\ae}r, M.~Slaney, and E.~Lalor, ``Decoding the auditory brain with canonical component analysis,'' \emph{NeuroImage}, vol. 172, pp. 206--16, May 2018.

\bibitem{thornton_robust_2022}
M.~Thornton, D.~Mandic, and T.~Reichenbach, ``Robust decoding of the speech envelope from {{EEG}} recordings through deep neural networks,'' \emph{J. Neural Eng.}, vol.~19, no.~4, Aug. 2022.

\bibitem{rotaru_what_2024}
I.~Rotaru, S.~Geirnaert, N.~Heintz, I.~{Van de Ryck}, A.~Bertrand, and T.~Francart, ``What are we really decoding? {{Unveiling}} biases in {{EEG-based}} decoding of the spatial focus of auditory attention,'' \emph{J. Neural Eng.}, vol.~21, no.~1, Feb. 2024.

\bibitem{puffay_relating_2023}
C.~Puffay, B.~Accou, L.~Bollens, M.~J. Monesi, J.~Vanthornhout, H.~{Van hamme}, and T.~Francart, ``Relating {{EEG}} to continuous speech using deep neural networks: A review,'' \emph{J. Neural Eng.}, vol.~20, no.~4, Aug. 2023.

\bibitem{szegedy_going_2015}
C.~Szegedy, W.~Liu, Y.~Jia, P.~Sermanet, S.~Reed, D.~Anguelov, D.~Erhan, V.~Vanhoucke, and A.~Rabinovich, ``Going {{Deeper With Convolutions}},'' in \emph{In: Proceedings of the {{IEEE Conference}} on {{Computer Vision}} and {{Pattern Recognition}}}, 2015, pp. 1--9.

\bibitem{zhang_modern_2021}
A.~Zhang, Z.~C. Lipton, M.~Li, and A.~J. Smola, ``Modern convolutional neural networks,'' in \emph{Dive into Deep Learning}.\hskip 1em plus 0.5em minus 0.4em\relax Cambridge University Press, 2021, pp. pp. 268--325.

\bibitem{he_deep_2016}
K.~He, X.~Zhang, S.~Ren, and J.~Sun, ``Deep {{Residual Learning}} for {{Image Recognition}},'' in \emph{Proceedings of the {{IEEE Conference}} on {{Computer Vision}} and {{Pattern Recognition}}}, 2016, pp. 770--78.

\bibitem{bai_empirical_2018a}
S.~Bai, J.~Z. Kolter, and V.~Koltun, ``An {{Empirical Evaluation}} of {{Generic Convolutional}} and {{Recurrent Networks}} for {{Sequence Modeling}},'' Apr. 2018.

\bibitem{lea_temporal_2017}
C.~Lea, M.~D. Flynn, R.~Vidal, A.~Reiter, and G.~D. Hager, ``Temporal {{Convolutional Networks}} for {{Action Segmentation}} and {{Detection}},'' in \emph{Proceedings of the {{IEEE Conference}} on {{Computer Vision}} and {{Pattern Recognition}}}, 2017, pp. 156--65.

\bibitem{aadnet_2025_github}
N.~D.~T. Nguyen, ``Aadnet: Auditory attention decoding neural network,'' \url{https://github.com/babibo180918/AADNet} (accessed 05. 08. 2025), 2025.

\bibitem{ingolfsson_eegtcnet_2020}
T.~M. Ingolfsson, M.~Hersche, X.~Wang, N.~Kobayashi, L.~Cavigelli, and L.~Benini, ``{{EEG-TCNet}}: {{An Accurate Temporal Convolutional Network}} for {{Embedded Motor-Imagery Brain}}--{{Machine Interfaces}},'' in \emph{2020 {{IEEE International Conference}} on {{Systems}}, {{Man}}, and {{Cybernetics}} ({{SMC}})}, Oct. 2020, pp. 2958--65.

\bibitem{chollet_xception_2017}
F.~Chollet, ``Xception: {{Deep Learning}} with {{Depthwise Separable Convolutions}},'' Apr. 2017.

\bibitem{das_auditory_2019}
N.~Das, T.~Francart, and A.~Bertrand, ``Auditory attention detection dataset kuleuven,'' \url{https://doi.org/10.5281/zenodo.4004271}, 2019, [dataset].

\bibitem{fuglsang_eeg_2018}
S.~A. Fuglsang, D.~D. Wong, and J.~Hjortkj{\ae}r, ``Eeg and audio dataset for auditory attention decoding,'' \url{https://doi.org/10.5281/zenodo.1199011}, 2018, [dataset].

\bibitem{azure_2023_framework}
{Microsoft Corporation}, ``Microsoft azure text-to-speech service,'' \url{https://azure.microsoft.com/en-us/products/ai-services/text-to-speech} (accessed 22. 03. 2023), 2023.

\bibitem{gramfort_meg_2013}
A.~Gramfort, M.~Luessi, E.~Larson, D.~A. Engemann, D.~Strohmeier, C.~Brodbeck, R.~Goj, M.~Jas, T.~Brooks, L.~Parkkonen, and M.~H{\"a}m{\"a}l{\"a}inen, ``{{MEG}} and {{EEG}} data analysis with {{MNE-Python}},'' \emph{Front. Neurosci.}, vol.~7, p. 267, Dec. 2013.

\bibitem{virtanen_scipy_2020}
P.~Virtanen, R.~Gommers, T.~E. Oliphant, M.~Haberland, T.~Reddy, D.~Cournapeau, E.~Burovski, P.~Peterson, W.~Weckesser, J.~Bright, S.~J. {van der Walt}, M.~Brett, J.~Wilson, K.~J. Millman, N.~Mayorov, A.~R.~J. Nelson, E.~Jones, R.~Kern, E.~Larson, C.~J. Carey, {\.I}.~Polat, Y.~Feng, E.~W. Moore, J.~VanderPlas, D.~Laxalde, J.~Perktold, R.~Cimrman, I.~Henriksen, E.~A. Quintero, C.~R. Harris, A.~M. Archibald, A.~H. Ribeiro, F.~Pedregosa, and P.~{van Mulbregt}, ``{{SciPy}} 1.0: Fundamental algorithms for scientific computing in {{Python}},'' \emph{Nat. Methods}, vol.~17, no.~3, pp. 261--72, Mar. 2020.

\bibitem{geirnaert_interpretable_2020}
S.~Geirnaert, T.~Francart, and A.~Bertrand, ``An {{Interpretable Performance Metric}} for {{Auditory Attention Decoding Algorithms}} in a {{Context}} of {{Neuro-Steered Gain Control}},'' \emph{IEEE Transactions on Neural Systems and Rehabilitation Engineering}, vol.~28, no.~1, pp. 307--17, Jan. 2020.

\bibitem{mesdtoolbox_2023_github}
S.~Geirnaert, D.~Fierberg, ``Mesd toolbox: Minimal expected switch duration for auditory attention decoding,'' \url{https://github.com/exporl/mesd-toolbox} (accessed 20. 12. 2024), 2023.

\bibitem{paszke_pytorch_2019a}
A.~Paszke, S.~Gross, F.~Massa, A.~Lerer, J.~Bradbury, G.~Chanan, T.~Killeen, Z.~Lin, N.~Gimelshein, L.~Antiga, A.~Desmaison, A.~K{\"o}pf, E.~Yang, Z.~DeVito, M.~Raison, A.~Tejani, S.~Chilamkurthy, B.~Steiner, L.~Fang, J.~Bai, and S.~Chintala, ``{{PyTorch}}: An imperative style, high-performance deep learning library,'' in \emph{Proceedings of the 33rd {{International Conference}} on {{Neural Information Processing Systems}}}, no. 721.\hskip 1em plus 0.5em minus 0.4em\relax Red Hook, NY, USA: Curran Associates Inc., Dec. 2019, pp. 8026--37.

\bibitem{harris_array_2020}
C.~R. Harris, K.~J. Millman, S.~J. {van der Walt}, R.~Gommers, P.~Virtanen, D.~Cournapeau, E.~Wieser, J.~Taylor, S.~Berg, N.~J. Smith, R.~Kern, M.~Picus, S.~Hoyer, M.~H. {van Kerkwijk}, M.~Brett, A.~Haldane, J.~F. {del R{\'i}o}, M.~Wiebe, P.~Peterson, P.~{G{\'e}rard-Marchant}, K.~Sheppard, T.~Reddy, W.~Weckesser, H.~Abbasi, C.~Gohlke, and T.~E. Oliphant, ``Array programming with {{NumPy}},'' \emph{Nature}, vol. 585, no. 7825, pp. 357--62, Sep. 2020.

\bibitem{mahjoory_convolutional_2024}
K.~Mahjoory, A.~Bahmer, and M.~J. Henry, ``Convolutional neural networks can identify brain interactions involved in decoding spatial auditory attention,'' \emph{PLOS Computational Biology}, vol.~20, no.~8, Aug. 2024.

\bibitem{chollet2017xception}
F.~Chollet, ``Xception: Deep learning with depthwise separable convolutions,'' in \emph{Proceedings of the IEEE Conference on Computer Vision and Pattern Recognition (CVPR)}, 2017, pp. 1251--58.

\bibitem{geirnaert_direct_2025a}
S.~Geirnaert, S.~L. Kappel, and P.~Kidmose, ``A direct comparison of simultaneously recorded scalp, around-ear and in-ear {{EEG}} for neural selective auditory attention decoding to speech,'' \emph{Sci Rep}, vol.~15, no.~1, p. 41655, Nov. 2025.

\bibitem{haufe_interpretation_2014}
S.~Haufe, F.~Meinecke, K.~G{\"o}rgen, S.~D{\"a}hne, J.-D. Haynes, B.~Blankertz, and F.~Bie{\ss}mann, ``On the interpretation of weight vectors of linear models in multivariate neuroimaging,'' \emph{NeuroImage}, vol.~87, pp. 96--110, Feb. 2014.

\bibitem{kubanek_tracking_2013}
J.~Kubanek, P.~Brunner, A.~Gunduz, D.~Poeppel, and G.~Schalk, ``The {{Tracking}} of {{Speech Envelope}} in the {{Human Cortex}},'' \emph{PLOS ONE}, vol.~8, no.~1, p. e53398, Jan. 2013.

\bibitem{vanderghinst_left_2016}
M.~Vander~Ghinst, M.~Bourguignon, M.~{Op de Beeck}, V.~Wens, B.~Marty, S.~Hassid, G.~Choufani, V.~Jousm{\"a}ki, R.~Hari, P.~Van~Bogaert, S.~Goldman, and X.~De~Ti{\`e}ge, ``Left {{Superior Temporal Gyrus Is Coupled}} to {{Attended Speech}} in a {{Cocktail-Party Auditory Scene}},'' \emph{J. Neurosci.}, vol.~36, no.~5, pp. 1596--1606, Feb. 2016.

\bibitem{tanveer_deep_2024}
M.~A. Tanveer, M.~A. Skoglund, B.~Bernhardsson, and E.~Alickovic, ``Deep learning-based auditory attention decoding in listeners with hearing impairment,'' \emph{J. Neural Eng.}, vol.~21, no.~3, May 2024.

\bibitem{straetmans_neural_2024}
L.~Straetmans, K.~Adiloglu, and S.~Debener, ``Neural speech tracking and auditory attention decoding in everyday life,'' \emph{Front. Hum. Neurosci.}, vol.~18, Nov. 2024.

\end{list}


\end{document}